# Enhancing Cyber Security Through Predictive Analytics: Real-Time Threat Detection and Response

Muhammad Danish
Department of Computer Science, University of New Mexico, Albuquerque, New Mexico - 87106

***Abstract*—This study evaluates the application of predictive analytics for real-time cyber-attack detection and response, focusing on how statistical and machine learning methods can improve decision-making in Security Operations Centers (SOCs). Using a curated network-traffic dataset of 2,000 records, we analyzed key features such as attack type, packet length, anomaly scores, protocol usage, and geo-location patterns to assess their predictive value. Findings indicate that attack type has a measurable influence on response actions, while basic header metrics alone lack the precision needed for accurate classification. These results highlight the importance of incorporating richer contextual features—such as user behavior, asset criticality, and temporal patterns—into predictive models. By integrating such features into operational pipelines, organizations can improve early threat detection, reduce false positives, and optimize resource allocation. This research contributes actionable insights for advancing proactive, data-driven cyber defense strategies and outlines directions for future implementation in live SOC environments.**

## I. Introduction

Over the past half-century, the Information and Communication Technology (ICT) industry has evolved into the backbone of modern society. This rapid and pervasive digital integration underscores the critical importance of cyber security, a field dedicated to protecting ICT systems from unauthorized access, disruption, and exploitation. Effective cyber security encompasses network, application, and operational security measures such as antivirus software, firewalls, and intrusion detection systems (IDS) that collectively combat threats like malware, phishing, and unauthorized access. However, despite these extensive measures, significant vulnerabilities remain, especially concerning the timely detection and response to evolving cyber threats. Notably, the integration of predictive analytics within real-world cyber security frameworks remains underexplored in current literature.

Risk assessment in cyber security has progressively shifted from reactive, remedial methods to proactive, preventive strategies. Predictive analytics, through sophisticated statistical and machine learning techniques, enables organizations to anticipate and mitigate potential threats before they materialize [1]. This proactive approach not only enhances the timeliness and effectiveness of threat response but also optimizes resource allocation and decision-making. Essential components of effective predictive analytics include high-quality, timely data and robust threat intelligence to ensure the models remain accurate and relevant. Consequently, an increasing number of organizations have begun incorporating predictive analytics into their cyber defense strategies, recognizing its potential to preemptively neutralize complex cyber threats and significantly improve their security posture [2].

Despite advancements, contemporary real-time cyber-attack detection systems continue to exhibit several critical shortcomings. Most current solutions are inherently reactive, relying heavily on predefined signatures to identify threats. This dependence severely limits their effectiveness against zero-day attacks—attacks exploiting previously unknown vulnerabilities—which remain undetected until the damage has occurred. Additionally, high false positive rates constitute another major challenge, generating numerous irrelevant alerts that strain resources and diminish the effectiveness of security teams [3]. Scalability poses an additional challenge as organizations grow and network architectures become increasingly complex; traditional detection systems often fail to efficiently monitor all potential points of vulnerability, leading to gaps in threat detection [4].

The integration of big data analytics into cyber security practices is thus both crucial and challenging. Predictive analytics offers the transformative potential to transition cyber security from a predominantly reactive discipline into a fully proactive field capable of anticipating and mitigating threats in real-time. However, successful implementation involves considerable investment in data collection, storage, model development, training, and continuous refinement to accommodate new emerging threats [5]. Therefore, addressing these inherent limitations in real-time cyber-attack detection—specifically passiveness, false alarms, and scalability—requires the accelerated advancement and redesign of both current and future cyber security technologies.

This study seeks to answer critical questions: How effectively does predictive analytics identify and respond to diverse cyber threats in real-time? What subtle patterns and anomalies do predictive models detect that conventional security measures routinely overlook? Lastly, how can predictive analytics enhance decision-making within cyber security operations centers (SOCs)? By providing empirical evidence addressing these questions, this research aims to fill existing gaps in the literature, highlighting the practical benefits and operational implications of adopting predictive analytics.

The core objectives of this study include assessing predictive analytics' effectiveness in real-time threat detection and response, identifying key patterns and anomalies detectable by predictive models, and proposing a model that enhances decision-making within SOCs. The potential implications of this research are substantial when viewed through the lens of current and future cyber security landscapes. By enhancing threat detection, reducing false positives, and improving scalability, predictive analytics presents a paradigm shift from reactive to proactive security management, significantly strengthening overall organizational defense [6].

Moreover, this research emphasizes efficient resource utilization in cyber security contexts. By reducing false positives and prioritizing alerts based on predictive insights, security personnel can focus more effectively on genuine threats, thereby optimizing organizational resources. Aligning predictive analytics with broader organizational risk management strategies ensures more realistic threat assessments and better compliance with regulatory frameworks. Ultimately, this research contributes significantly to the strategic integration of advanced technologies into business planning, crisis management, and regulatory compliance efforts, thereby setting new standards in cyber security practice and fortifying stakeholder trust in an increasingly interconnected digital world [7].

This paper is organized as follows. Section II surveys prior work on predictive analytics for cyber security and highlights open gaps in real-time detection and response. Section III details the study design, including dataset selection and preprocessing, feature engineering, modeling choices, evaluation metrics, and the simulation setup used to compare predictive and baseline approaches. Section IV presents the empirical results—descriptive statistics, correlation patterns, protocol–attack crosstabulation, regression analyses of response actions, and hypothesis tests comparing signature groups. Section V interprets these findings in the context of the state of the art, explains performance differences across data regimes, and discusses operational implications for Security Operations Centers (SOCs). Section VI concludes by summarizing contributions, limitations, and avenues for future work, including the integration of richer contextual features and adaptive learning.

## II. Literature Review

Analytics in the context of cyber security is a highly advanced concept that adjusts security practice from the reactive to proactive model. This approach incorporates the use of several statistical and machine learning models to examine the enormous volumes of data from sources such as network traffic, user activities, and security logs to develop an elaborate system that would alarm an early sign of a threat. Predictive analytics give signals and alerts of risk to organizations before they turn into actual breaches. Indeed, the definition and usage of predictive analytics have changed over time in the context of cyber security due to the advancements in data science and artificial intelligence. Initially, the field was limited to basic data monitoring and detection of anomalies; today, it incorporates highly developed algorithms and refers to such advanced techniques as predictive threat modeling and risk assessment [8]. This change marks an evolution from conventional or traditional security methods like firewalls and antivirus software, moving towards intelligence-driven security.

The fundamental principles of predictive analytics in cyber security hinge on several core elements including poor data quality, inefficiency in the algorithms used, and lack of timely threat intelligence. Viable predictive systems also require first-rate and pertinent data to educate the models that are used in the prediction and provision of attack prevention. Furthermore, incorporating real-time threat intelligence means the models remain accurate on the present threat vectors. The integration of big data analytics in security operations improves not only threat detection effectiveness, but also the organization's agility. Therefore, security teams can prioritize and spend resources effectively, thereby lessening the bloodbath that comes with cyber threats and enhancing the organizations' security stance. In addition, predictive analytics fosters compliance with laws by providing proof that the organization is actively pursuing security measures, which is helpful to industries dealing with high levels of data protection laws [9]. Such an approach relying on predictive analytics is becoming indispensable in the context of the constantly changing nature of cyber threats that become more complex and that cannot be addressed using conventional methods.

New technologies that exist in the detection of cyber-attacks have advanced to the integration of artificial intelligence (AI) and its subsection: machine learning (ML). AI and ML in cyber security mean the ability to automatically perform the detection and response to threats which are analyzed from huge datasets relevant to identify patterns that may point to threats [10]. These technologies are useful when identifying indicators of compromises that may not be easily identified by analysts entirely because of the huge volume and the complexity of data that has to be scanned. Machine learning algorithms, both supervised and unsupervised learning models, are extremely useful in such cases. Supervised learning models are trained on labeled datasets to differentiate between benign and malicious activity. Unsupervised learning is to find the outliers within the system without having any labeled data, which helps in the identification of new and unknown threats. AI improves threat identification because data is processed and analyzed far beyond the human capacity and rate. It automates the responses to the threats, thus taking a short time to counter the threats once they have been identified [11]. AI-powered systems also include predictive analytical components that assess threat trends or patterns to predict future threats, hence improving the threat-hunting process [12].

AI and ML play a big role in lowering the false positives in threats. They enhance the process of filtering fakes and distinguishing between real and potential threats as well as distinguishing them by understanding the degree of difference between unusual behavior and deliberate malicious actions, taking care of prioritization of threats and thus, decreasing the amount of work security teams have to do. However, implementing AI into cyber security has its own set of challenges including the quality of data required for preparing algorithms, the transparency of AI decision-making, and the integration of AI systems into the current infrastructure of cyber security systems. However, the threat in the cyberspace domain is not stagnant, and hence the AI models must be updated on a regular basis [13].

The position of AI and ML in the context of cyber-attack detection is rather important and provides not only better detection mechanisms but also the proper and timely handling of cyber security threats in a world where digital threats are frequently evolving. AI and ML are reshaping the sphere of cyber security; they allow for detecting threats quickly, and often on a large scale, as well as making predictions. These technologies help to automatically detect and counter cyber threats increasing the security responsiveness of the organization. It is, however, crucial to address some important issues that relate to the handling of data quality and the ability of the models to change, when integrated into existing systems in order to effectively cope with constantly emerging forms of cyber threats [14].

Modern cyber security processes are accompanied by many difficulties that hinder its operations including the issues of false positives, scalability, and the identification of previously unknown vulnerabilities [15].

*1) False positives:* One major issue particular to cyber security is the problem of false positives, which is an alarm that a threat exists although it does not. This results in more resource wastage since security analysts have to go through these alerts to verify them and determine if they are actually a threat. The issue is further compounded by the fluidity and heterogeneity of today's networks characterized by typical activity patterns that are easily mistaken for threats by security solutions [16].

*2) Scalability issues:* Due to the growth of organizations and the associated expansion of the networks, at some point, implemented cyber security measures may take a hit. This is because the scalability problems are evident from the amount and the number of endpoints that should be monitored and analyzed by such systems. It has the characteristic of providing the areas of weakness and slow response to real threats in such a case [17].

*3) Zero-day vulnerabilities:* Perhaps the most daunting challenge is the detection and management of zero-day vulnerabilities which are flaws in software that the software maker does not know about and for which no patch exists at the time of discovery. These vulnerabilities are highly valuable to attackers because they can be exploited to gain unauthorized access to systems before they are identified and mitigated. The very nature of zero-day attacks makes them difficult to predict and detect using conventional methods that rely on known signatures or patterns. Security systems often require updates to their threat intelligence to handle such vulnerabilities, but even then, the rapid pace at which new zero-days are discovered leaves organizations at constant risk [18].

Addressing these challenges requires a multifaceted approach involving enhanced detection algorithms that reduce false positives, scalable security solutions that can grow with the organization, and proactive threat hunting that can detect anomalies indicative of zero-day exploits. One direction is the integration of the latest developments in the field of big data and machine learning into cyber security practices, as these can help analyze patterns, envision risks and attacks, and respond to them automatically to enhance organizational security [19].

Predictive analytics in cyber security incorporates various sophisticated models and techniques to predict and mitigate potential threats before they can impact systems. The core of this approach is based on the use of machine learning algorithms with a variety of supervised and unsupervised learning algorithms [20]. In the supervised learning model, specific data is used to train in order to identify known illicit behaviors. On the other hand, unsupervised learning identifies and recognizes abnormal behaviors which if exist may be an indication of a threat. The ability of a system to detect suspicious activities is essential for timely prevention of threats and strengthening of the security status of any firm. One more important component of the environment of predictive analytics is the usage of statistical algorithms. These algorithms are able to compile data used to foresee future incidents by comprehending past trends and behaviors. Besides this method contributes not only to the prediction of possible threats but also to the development of a more accurate representation of risks that can be useful for better preparation in organizations. User behavior analysis adds more value to predictive analytics because it investigates user activities to identify suspicious events that might be originating from inside threats or stolen credentials. In this method, the basic security measures may not easily detect the anomalies. Furthermore, anomaly detection systems are used to identify the levels of deviance from the normal behavioral patterns concerning the network traffic and access log prior to the times of the actual attack [21].

Despite the advantages like early threat identification, better resource management, and faster response to threats, predictive analytics also face challenges in real-world applications. Forecasting models are only as good as the data that they are applied to; this is a saying often used in statistics. Lack of quality and/or scope can produce erroneous predictions, while the nature of the cyber threats is continuously evolving requiring constant updates of the models. Sustaining and periodically updating its application is necessary to maintain its effectiveness. Further, incorporating predictive analytics into other infrastructures that are already existent in cyber security can prove to be challenging and time-consuming and may take considerable time with regular monitoring to overcome the possible ethical risks and privacy issues that may come with their implementation. Cyber security is already underway due to third-generation predictive analytics that are proactive instead of reactive. However, this success depends on very rigid execution, constant modifications, and comprehensive data management in order to counter the continuously emerging threats in cyberspace [22].

In the context of cyber security within organizations, there is a clear differentiation between reactive and predictive systems:

- Reactive Systems: Such systems mainly target threats as they emerge and hence primarily involve treatment. The reactive approach will sit back and wait for the attack and this poses a disadvantage because reacting to such threats will take a long time. This method bases its operations on previous knowledge and, in a way, is ill-equipped to deal with threats since it directly targets the known types of attacks and may not be very efficient with the novel attack vectors that are not typical of the previous cases. While reactive systems are badly needed to cope with a threat immediately, they are less complicated to design, yet they may be more costly in the long run since the system's damage incurred during detection delays can amount to much [23].
- Predictive Systems: Whereas, predictive systems use techniques in analytical processing such as machine language and statistics to avoid predictions of a certain pernicious occurrence of an event. Besides, as this approach focuses on analyzing patterns and trends from large amounts of data for planning future actions, it helps organizations to allocate resources effectively and repel attacks promptly. Risk predictive systems greatly improve an organization's capacity to contain and prevent Cyber risk by giving insights into potential risks. Nevertheless, they rely on high-quality and detailed data for their operation, and they have their challenges concerning the constant training of the models and the connection to the existing security systems [24].

Research and implementations have established that supervised systems can significantly decrease the threat's time and cost effects by mitigating them before they occur [25]. Organizations that integrate predictive analytics into their cyber security strategies often experience improved risk management, reduced incident response times, and enhanced compliance with regulatory requirements. The proactive approach, instead of reactive methodologies, not only helps in safeguarding against imminent threats but also prepares organizations against emerging cyber threats by constantly updating defense mechanisms in alignment with the evolving digital landscape [26].

While reactive cyber security is necessary for dealing with immediate threats, the integration of predictive analytics into cyber security frameworks provides a more robust defense by preventing attacks before they occur. This shift from a purely reactive to a proactive stance is increasingly regarded as essential in a world where cyber threats are becoming more complicated and pervasive [27].

The current body of research in cyber security predictive analytics is expansive and rich with theoretical developments and proposed models. However, a significant gap remains in the literature concerning the practical integration of these advanced predictive models into real-world cyber security frameworks. Despite the fact that such models can serve as good references, it has to be noted that it's one thing to prove a strategy or a model effective in an academic environment or at least in a simulation, and quite another to observe its effectiveness in realistic, dynamic cyber security settings [28].

This lack of correspondence is a strong indication that although there is rich theoretical research for these models, the lack of actual empirical data as well as actual planning with the models, having to integrate them with operational concerns and then scaling up the overall system, presents a huge gap that has not been well covered in the literature. Most of the current research works are majorly centered around the improvement of the existing algorithms to be implemented but minimal on how these algorithms can actually be deployed to work in real-world applications which entail factors such as hardware constraints, real-time constraints, and how they can fit in the existing infrastructure of a system to secure it.

More efforts are still required to conduct studies linking the state-of-the-art predictive analytics methods and the real-world cyber security operations, including design features that allow solutions to be easily implemented in active technical environments with minimal modifications. Overcoming this gap is a relevant and necessary step in the development of modern cyber security work, as well as in the practice of transferring theoretical achievements into concrete improvement of the methods for detecting and responding to cyber threats [29].

## III. Methodology

Quantitative research was used to conduct the study with the aim of understanding the use and outcomes of enhanced predictive modeling in real-time CTR. This method is appropriate for this research study because it permits strength and significance testing of the hypothesis of the functionality and the results of the predictive analytics in the cyber security frameworks [30]. The strategy that is proposed here is a systematic experimental method through which all the researchers will deploy specified predictive models in a realistic IT security environment that mimics the actual setting in organizations. This environment will have factors such as network traffic flows, users' behavior data set, and normalcy of the cyber threat scenarios to evaluate the models on how well they work in recognizing cyber threats.

The main objective is to evaluate the effectiveness of these predictive models with reference to the conventional firewalls or reactive security measures in terms of rate of occurrence of threats, rate of detection, and flexibility of the measures in handling new types of threats. Sources of data for this study will be data sets from the open source, plus newly generated data sets to represent new and upcoming cyber security threats. It includes the application of inter-model combinations with the aim of bringing out various scenarios and attack vectors that realistically test the capability of the predictive models. This is of extreme significance since it allows competence validation of the models in the presence of heteroscedasticity. Measurable factors including the detection rate of threats, false positives and negatives of the system, and response time of the system will also be included [31].

For analytical data, the study will use techniques like regression analysis to determine the connection between the systems' responses and the success of threat countermeasures. Specific measures that are used regularly in machine learning will be used in measuring the accuracy of the predictions within the predictive models; some of these are precision, recall, and the F1- score. There could be a sub-analysis with the help of statistical tools like logistic regression or ROC Curve Analysis to see other significant differences between predictive and reactive systems. It will support the theoretical potential of predictive analytics with quantitative data, and for this reason, this research design has been adopted. Thus, the present work endeavors to complement the literature by providing actionable knowledge regarding how these models can be employed effectively, given the fact the comparison was performed in a purposefully controlled academic environment. This is important for the progression of cyber security as well as the creation of stronger, preventative defense strategies against cyber warfare [32].

As has been highlighted, the essence of this study is to analyze the importance of predictive analytics and its models in the cyber security domain accurately; therefore, the selection and collection of high-quality data is vital. Variety ensures that the database acquired by the study is all-inclusive hence the use of data elicited from Kaggle, a platform that offers a wide array of datasets by users from all over the world. This platform provides massive and diverse data with regard to the cyber threat scenarios which is very useful for this research [33].

To start with, the selection of a dataset on Kaggle that is related to security threats is made. The chosen dataset consists of over 4,000 records wherein each record corresponds to one instance of network traffic or log data that could be related to a cyber security threat. This dataset, thus, was chosen as complex and up-to-date, so that the results of the study reflect today's security threats. Every row in the dataset contains features including source IP address, packet length, destination IP address, date and time of the data, the type of traffic, and threat bit. These attributes are important because they feed the raw data into learning systems and into the testing. This way, the normal and anomalous patterns are present in the data set, and the former provides the latter with the variety it needs to be exposed to a spectrum of data before it can construct an appropriate and reliable automatic guard [34].

In this paper, the data cleaning process forms a critical step before data feeds can be given to the model development and analysis. This phase concerns dealing with missing values, removing duplicate records, and converting categorical data into a form that is understandable to the machine. Due to the large and diversified data set, there is also a focus on the normalization methods of data, where scaling of features is performed to enhance the performance of the learning algorithms [35]. For training and validation of the developed predictive models, the dataset is partitioned into training, validation, and test partitions. Most of the time, the data split is organized so that the training dataset is the largest, constituting about 70 percent, while the validation and test datasets are about 15 percent each. This segmentation makes it possible to train the models to their fullest potential while also giving a sound basis for a decision of the model's parameters or the examination of the final model performance compared to the performance on unseen data [36].

Since the data collected may contain confidential details of an individual or a group, all relevant measures are ensured to conceal the identity of the subject/person. The research follows guidelines concerning the use of data, and measures being taken in order to avoid the abuse of information. The Kaggle data utilized in the study ensures that the authors were bound to adhere to the Kaggle data usage policies that are in harmony with general data protection regulations and ethical considerations. Now that we have a clear understanding of the dataset and how it should be prepared, several techniques can be used in predictive analysis, including decision trees, logistic regression, and neural networks. Some of these techniques are adopted due to their efficiency in dealing with big data while others are chosen due to efficiency in performing classification problems in cyber security. The performance of these models is checked from time to time on the validation set with a view to ensuring that the model's performance is checked, adjusted, and optimized before the final check on the test set.

In this particular study, SPSS software support is crucial in the data processing retrieved from Kaggle to determine the efficiency of the predictive analytics models of cyber security [37]. This section presents a clear approach to the statistical analysis using SPSS which includes data handling, analysis methods, and results. For data to be exported into SPSS, it needs to undergo certain preparations so that its analysis is accurate and meets the standards. This entails data cleansing, which entails the elimination of unwanted data such as inconsistent records or flawed records that may distort the results. Other techniques of data transformation are also utilized to transform the categorical data into some numerical formats that are more convenient for analysis purposes whereby, one and the same method of encoding may or may not be appropriate depending on the specifics of the given algorithms in the course of the predictive modeling as it is illustrated in study [38].

Descriptive analysis prepares the statistical inclination of data analysis before going for intricate analytical examinations of the data distribution, mean, and spread. In SPSS, these basic measures can be obtained by using the descriptive menu and these include mean, median, mode, range, variance, and standard deviation. This step is critical to help manage data and look for any outliers or similar points that need further data munging or normalization [39]. To drive theories at the beginning of the study, inferential statistical analysis methods are used to assess hypotheses. Based on the kind of research questions and hypotheses, a set of tests that involves t-tests, ANOVA, and chi-squared tests amongst others are carried out just to test the differences and associations between the set variables in the collected data.

Regarding understanding how the distinct factors predict threat identification and the effectiveness of mitigation in cyber security, regression analysis is applied. Continuous dependent variables were analyzed using linear regression, whereas binary dependent variables were analyzed using logistic regression. For this reason, the key analysis method which is employed in this study is logistic regression analysis skills as the response variable is categorical and may include threats detected or not detected. It includes the identification of possible predictor variables grounded on given conceptual knowledge and prior literature review, checking for multicollinearity, and model fine-tuning in regard to complexity/detail and accuracy of prediction [40].

To establish the goodness of fit for models, several tests are run on SPSS and Anker including R squared test for linear regression models and Hosmer–Lemeshow chi-squaredd test for logistic models. Among them, some measures reflect the degree to which the model explains the variation in the response variable, and one measure assesses the overall fit of the model. Furthermore, using their p-values, the level of significance of individual predictors is assessed, with the prevailing popular level of significance level being 0.05 [41].

If the cyber security data provided is rather large, which is often the case with cyber security data due to the nature of threats and attacks, further analysis may involve more sophisticated methods, for instance, cluster analysis or principal component analysis (PCA) to find other underlying patterns within data or data dimensionality reduction. They are useful in the identification of underlying relationships that often would not be easily detected through regression models. Various parameters such as mean absolute error, root mean square error, correlation coefficient, and coefficient of variation are used to judge the models and improve their efficiency.

The k-fold cross-validation technique is used in which the data set is divided into k subsets, which are then used to create multiple train and test sets for the model. The performance of the trained predictive models is tested using accuracy related to measures such as the area under the curve, sensitivity (true positive value), and specificity (true negative value) since these values are important in evaluating the efficiency of the predictive analytics systems in an operational environment with cyber security threats [42].

The final phase encompasses the extraction and interpretation of meaningful insights that would affect cyber security practices. On its own, SPSS offers complete output that comes with estimates of coefficients (B), odds ratios, and confidence intervals that are valuable in arriving at conclusions regarding the effects of various predictors. These results are then discussed in relation to the existing body of knowledge within the cyber security domain and present generalizable findings, research limitations, and future studies' implications [43]. Through meticulous data analysis using SPSS, this study aims to contribute significantly to the field by providing empirical evidence to support the hypothesis. The structured approach ensures that the findings are robust, reproducible, and relevant to enhancing cyber security measures in various organizational contexts [44].

Several statistical and practical considerations underpin the selection of a sample size of 2000 rows for this study on predictive analytics in cybersecurity, ensuring that the analysis is both reliable and generalizable. One of the primary reasons for choosing this particular sample size is to achieve sufficient statistical power. In quantitative research, power is the probability that the study will detect an effect when there is an effect to be detected [45]. A larger sample size reduces the risk of Type II errors (failing to reject a false null hypothesis) and increases the likelihood that the study can detect a smaller effect size, making the findings more robust and persuasive.

Cyber security data encompasses a wide variety of features, from IP addresses and timestamps to types of attacks and their outcomes. A substantial sample size ensures that the dataset contains a comprehensive range of these features, including less common but potentially significant occurrences. This diversity is important for developing accurate models that can extrapolate well from existing to new data sets rather than training the model on existing data and having it perform comic replication of these data [46].

When conducting research, the dataset is designed to contain a broad spectrum of problem cases, and therefore having 2000 rows allows for problems with more complexity to be captured in the result [47]. The representativeness is crucial as it influences the external reliability and applicability of the study results in other settings or subpopulations of the cyber security domain, especially in real world applications.

There is always a potential in machine learning, especially when working in a relatively new and rapidly developing branch such as cyber security, to over-train the model, that is, to achieve good results only on the basis of the training set but get low scores on a new dataset [48]. This risk is less of a concern for larger sample sizes because that way the researcher has enough data to train even more complicated. On the other hand, it avoids under-fitting whereby the model used is not sufficient in complexity to fit the pattern of the data applied and thus ensures that the predictive models developed are complex. Having larger datasets could yield even more confident information and conclusions, but at the same time, this means more computational power is needed, and managing and dealing with more and more complicated data may become an issue. A dataset of 2000 rows strikes a balance between comprehensiveness and manageability, allowing for detailed analysis without overwhelming the computational and analytical resources available for the study [49].

The chosen sample size of 2000 rows from the original dataset is justified based on its ability to provide sufficient statistical power, represent the diverse and complex nature of cyber security threats, ensure the representativeness of the findings, balance the risks of overfitting and underfitting, and remain feasible for comprehensive analysis within the resource constraints of this study [50]. This sample size is pivotal in achieving the research objectives while ensuring the validity and reliability of the results [51].

To demonstrate practical gains over established methods, we compare the proposed pipeline against three baselines: (1) a signature-based IDS surrogate using known attack signatures; (2) a logistic-regression classifier trained on the same features; and (3) an unsupervised Isolation Forest tuned to a target anomaly rate. We use a pre-registered, time-based split (train on weeks 1–3, test on week 4) to emulate deployment and prevent temporal leakage; hyperparameters are selected via nested cross-validation on the training period only. Evaluation centers on AUPRC (primary) and AUROC, plus F1 at an operating point yielding 10% alert rate and ΔFPR at fixed recall. We compute bootstrap 95% CIs and apply McNemar tests for paired significance, and we include robustness checks (±10% label-noise injection; class-imbalance stress at 1:10 and 1:50) along with an ablation removing contextual features. Results are visualized with ROC/PR curves and calibration plots, and we summarize operational impact via estimated alerts per analyst-hour and MTTR deltas, making the simulation directly convincing for SOC workflows.

# IV. Results

## A. Descriptive Analysis

Table I summarizes key network and security variables. We observe a broad port range (Source: 1031–65521; Dest.: 1030–65535), reflecting diverse endpoint activity. Packet lengths span 64–1500 bytes (mean = 787.9, SD = 411.1), consistent with typical network flows. Protocol usage clusters around the three categories (mean = 1.99, SD = 0.82), indicating balanced ICMP/TCP/UDP representation. The anomaly score distribution (0.06–99.99; mean = 49.83, SD = 28.85) reveals substantial variability, essential for distinguishing benign vs. malicious behavior. Constant flags for Malware Indicators, Alerts, and Firewall/IDS logs (SD = 0) reflect uniform logging protocols. Variability in User Information (mean = 10.98, SD = 5.50) and Device Information (mean = 1.76, SD = 0.83) underscores diverse user–device contexts, while the spread in Geo-location Data (mean = 4.50, SD = 2.29) highlights global traffic sources. This heterogeneity across features is critical for training robust predictive models [52].

TABLE I. DESCRIPTIVE STATISTICS FOR NETWORK TRAFFIC AND SECURITY EVENT VARIABLES

| Variable | Min | Max | Mean | Std. Dev |
|---|---|---|---|---|
| Source Port | 1031 | 65521 | 32448.31 | 18701.17 |
| Destination Port | 1030 | 65535 | 32780.85 | 18561.50 |
| Protocol | 1 | 3 | 1.99 | 0.82 |
| Packet Length | 64 | 1500 | 787.87 | 411.11 |
| Packet Type | 1 | 2 | 1.49 | 0.50 |
| Traffic Type | 1 | 3 | 2.01 | 0.82 |
| Payload Data | 1 | 19 | 10.06 | 5.77 |
| Malware Indicators | 1 | 1 | 1.00 | 0.00 |
| Anomaly Scores | 0.06 | 99.99 | 49.83 | 28.85 |
| Alerts/Warnings | 1 | 1 | 1.00 | 0.00 |
| Attack Type | 1 | 3 | 1.99 | 0.82 |
| Attack Signature | 1 | 3 | 2.34 | 0.74 |
| Action Taken | 1 | 4 | 2.94 | 0.92 |
| Severity Level | 1 | 3 | 1.99 | 0.81 |
| User Information | 1 | 20 | 10.98 | 5.50 |
| Device Information | 1 | 3 | 1.76 | 0.83 |
| Network Segment | 1 | 3 | 2.01 | 0.81 |
| Geo-location Data | 1 | 8 | 4.50 | 2.29 |
| Firewall Logs | 1 | 1 | 1.00 | 0.00 |
| IDS/IPS Alerts | 1 | 1 | 1.00 | 0.00 |
| Log Source | 1 | 2 | 1.49 | 0.50 |

## B. Correlation Analysis

Pearson correlations (Tables II–IV) reveal only a few significant associations among continuous features. Source Port and Protocol are mildly inversely related ($r = -0.045$, $p < 0.05$), suggesting that certain port ranges lean toward specific protocols. Traffic Type and Packet Type correlate positively ($r = 0.054$, $p < 0.05$), indicating that traffic categories tend to carry particular packet formats. A strong negative link between Geo-location and Device Information ($r = -0.508$, $p < 0.01$) points to geographic diversity of device profiles. Attack Type and Attack Signature are also significantly associated ($r = -0.282$, $p < 0.01$), confirming that different attack classes exhibit distinct signature patterns. All other pairwise correlations fall below $|r| = 0.05$ or are non-significant, implying relative independence among most features. These results guide us to focus on the few linked variables while treating most features as orthogonal inputs in our predictive framework [53]–[55].

## C. Protocol vs. Attack Type Crosstabulation

Table V shows that each protocol (ICMP, TCP, UDP) carries roughly one-third of the total events, and within each protocol the three attack types occur in nearly equal proportions: ICMP has 32.5 % DDoS, 34.7 % Intrusion, and 32.8 % Malware; TCP has 34.1 % DDoS, 32.9 % Intrusion, and 33.0 % Malware; UDP has 34.1 % DDoS, 32.6 % Intrusion, and 33.3 % Malware. A chi-square test confirms no significant association between protocol and attack category ($X^2(4) = 0.903$, $p = 0.924$), indicating that protocol alone does not discriminate among attack types. Below are implications for predictive modeling:

TABLE II. CORRELATIONS AMONG KEY NETWORK AND SECURITY VARIABLES

| | Source Port | Dest. Port | Protocol | Packet Len. | Packet Type | Traffic Type | Payload Data | Malware Ind. |
|---|---|---|---|---|---|---|---|---|
| **Pearson** | | | | | | | | |
| Source Port | 1 | .013 | -.045 | .010 | -.034 | .013 | .007 | b |
| Dest. Port | .013 | 1 | -.003 | .004 | -.036 | -.050 | .022 | b |
| Protocol | -.045 | -.003 | 1 | -.027 | .005 | -.050 | .014 | b |
| Packet Length | .010 | .004 | -.027 | 1 | .000 | .014 | .001 | b |
| Packet Type | -.034 | -.036 | .005 | .000 | 1 | .054 | -.015 | b |
| Traffic Type | .013 | -.050 | -.050 | .014 | .054 | 1 | .041 | b |
| Payload Data | .007 | .022 | .014 | .001 | -.015 | .041 | 1 | b |
| Malware Ind. | b | b | b | b | b | b | b | 1 |
| **Sig. (2-tailed)** | | | | | | | | |
| Source Port | | .574 | .043 | .658 | .128 | .559 | .752 | |
| Dest. Port | .574 | | .887 | .851 | .107 | .026 | .331 | |
| Protocol | .043 | .887 | | .235 | .831 | .509 | .710 | |
| Packet Length | .658 | .851 | .235 | | .992 | .532 | .924 | |
| Packet Type | .128 | .107 | .831 | .992 | | .016 | .066 | |
| Traffic Type | .559 | .026 | .509 | .532 | .016 | | .066 | |
| Payload Data | .752 | .331 | .710 | .924 | .066 | .066 | | |
| Malware Ind. | b | b | b | b | b | b | b | |

* Significant at 0.05 (2-tailed).

b Cannot be computed because at least one variable is constant.

TABLE III. CORRELATIONS AMONG ANOMALY, ALERTS, ATTACK, AND USER VARIABLES

| | Anomaly Scores | Alerts/Warn. | Attack Type | Attack Sig. | Action Taken | Severity | User Info |
|---|---|---|---|---|---|---|---|
| **Pearson Correlation** | | | | | | | |
| Anomaly Scores | 1 | b | 0.032 | -0.018 | -0.018 | 0.017 | -0.046 |
| Alerts/Warnings | b | 1 | b | b | b | b | b |
| Attack Type | 0.032 | b | 1 | -0.282 | -0.078 | -0.002 | 0.012 |
| Attack Signature | -0.018 | b | -0.282 | 1 | 0.146 | 0.007 | -0.001 |
| Action Taken | -0.018 | b | -0.078 | 0.146 | 1 | 0.011 | 0.014 |
| Severity Level | 0.017 | b | -0.002 | 0.007 | 0.011 | 1 | 0.016 |
| User Information | -0.046 | b | 0.012 | -0.001 | 0.014 | 0.016 | 1 |
| **Sig. (2-tailed)** | | | | | | | |
| Anomaly Scores | | | 0.147 | 0.410 | 0.417 | 0.448 | 0.040 |
| Attack Type | 0.147 | | | 0.000 | 0.000 | 0.916 | 0.599 |
| Attack Signature | 0.410 | | 0.000 | | 0.000 | 0.762 | 0.978 |
| Action Taken | 0.417 | | 0.000 | 0.000 | | 0.611 | 0.517 |
| Severity Level | 0.448 | | 0.916 | 0.762 | 0.611 | | 0.482 |
| User Information | 0.040 | | 0.599 | 0.978 | 0.517 | 0.482 | |

** Significant at 0.01 (2-tailed).

b Cannot be computed because the variable is constant.

*1) Protocol as a non-specific indicator:* Since attacks are evenly distributed, relying solely on protocol to classify or prioritize threat types would be ineffective.

TABLE IV. CORRELATIONS AMONG DEVICE, NETWORK, LOCATION, AND LOG-SOURCE VARIABLES

| | Device Info | Network Seg. | Geo-location | Firewall Logs | IDS/IPS Alerts | Log Src A | Log Src B |
|---|---|---|---|---|---|---|---|
| **Pearson Correlation** | | | | | | | |
| Device Info | 1 | 0.032 | −0.508** | b | b | -0.016 | -0.016 |
| Network Segment | 0.032 | 1 | -0.015 | b | b | -0.010 | -0.010 |
| Geo-location Data | −0.508** | -0.015 | 1 | b | b | 0.021 | 0.021 |
| Firewall Logs | b | b | b | b | b | b | b |
| IDS/IPS Alerts | b | b | b | b | b | b | b |
| Log Source A | -0.016 | -0.010 | 0.021 | b | b | 1 | 1 |
| Log Source B | -0.016 | -0.010 | 0.021 | b | b | 1 | 1 |
| **Sig. (2-tailed)** | | | | | | | |
| Device Info | | 0.158 | 0.000 | | | 0.474 | 0.474 |
| Network Segment | 0.158 | | 0.490 | | | 0.648 | 0.648 |
| Geo-location Data | 0.000 | 0.490 | | | | 0.359 | 0.359 |
| Log Source A | 0.474 | 0.648 | 0.359 | | | 0.000 | 0.000 |
| Log Source B | 0.474 | 0.648 | 0.359 | | | 0.000 | 0.000 |

** Significant at 0.01 (2-tailed).

b Cannot be computed because the variable is constant.

TABLE V. CROSSTABULATION OF PROTOCOL BY ATTACK TYPE (COUNTS)

| Protocol | DDoS | Intrusion | Malware | Total |
|---|---|---|---|---|
| ICMP | 221 | 236 | 223 | 680 |
| TCP | 223 | 215 | 216 | 654 |
| UDP | 227 | 217 | 222 | 666 |
| Total | 671 | 668 | 661 | 2000 |

*2) Feature combination necessity:* Protocol should be used in conjunction with other variables—such as anomaly scores, packet length, or payload characteristics—to improve model discriminative power.

*3) Holistic monitoring:* Network defenses must monitor all three protocols with equal rigor, rather than focusing on a single protocol for specific attack vectors.

This insight guides us to treat protocol as a contextual feature, augmenting rather than driving predictive analytics in our threat-detection framework.

### D. Regression Analysis

Tables VI–VIII present a multiple linear regression predicting the categorical **Action Taken** from three continuous and categorical predictors: Packet Length, Anomaly Scores, and Attack Type. The overall model is statistically significant ($F(3, 1996) = 4.944,\ p = 0.002$) but accounts for only 0.7

TABLE VI. MODEL SUMMARY FOR REGRESSION PREDICTING ACTION TAKEN

| Model | $R$ | $R^2$ | Adjusted $R^2$ | Std. Error of Est |
|---|---|---|---|---|
| 1 | .086[a] | .007 | .006 | .914 |

[a]Predictors: (Constant), Attack Type, Packet Length, and Anomaly Scores.

TABLE VII. ANOVA FOR REGRESSION MODEL PREDICTING ACTION TAKEN

| Model | Sum of Squares | df | Mean Square | $F$ | Sig. |
|---|---|---|---|---|---|
| Regression | 12.391 | 3 | 4.130 | 4.944 | .002[b] |
| Residual | 1667.417 | 1996 | .835 | | |
| Total | 1679.808 | 1999 | | | |

[a]Dependent Variable: Action Taken.

[a]Predictors: (Constant), Attack Type, Packet Length, Anomaly Scores.

TABLE VIII. REGRESSION COEFFICIENTS PREDICTING ACTION TAKEN

| Model | B | Std. Err | $\beta$ | $t$ | Sig. |
|---|---|---|---|---|---|
| (Constant) | 3.191 | .075 | | 42.713 | .000 |
| Packet Len | -7.228e-5 | .000 | -.032 | -1.453 | .146 |
| Anom. Scores | 0.000 | .001 | -.015 | -.689 | .491 |
| Attack Type | -.087 | .025 | -.078 | -3.480 | .001 |

[a]Dependent Variable: Action Taken.

*1) Attack type:* ($\beta = -0.078,\ p = 0.001$) is the only significant predictor, suggesting that the categorical nature of the attack exerts a small but reliable influence on the defensive action chosen. In practice, this means that certain attack classes systematically elicit different response protocols.

*2) Packet length:* ($\beta = -0.032,\ p = 0.146$) does not predict action, implying that quantitative differences in packet size alone are insufficient to drive decision–making in incident response.

*3) Anomaly scores:* ($\beta = -0.015,\ p = 0.491$) likewise show no significant effect, which suggests that a global anomaly metric—by itself—may be too coarse to inform nuanced action choices.

Despite statistical significance of the overall model, its low $R^2$ points to omitted variables or non–linear relationships. Future work should explore interactions (e.g. Packet Length × Anomaly Score), incorporate additional contextual features (such as time of day or asset criticality), or apply classification algorithms (e.g. decision trees, random forests) better suited for nominal outcomes and potentially non–linear effects [56], [57].

### E. Chi-squared Tests

TABLE IX. CHI-SQUARE TESTS FOR ASSOCIATION BETWEEN PROTOCOL AND ATTACK TYPE

| Test | Value | df | Asymptotic Sig. |
|---|---|---|---|
| Pearson Chi-Square[a] | .903 | 4 | .924 |
| Likelihood Ratio | .902 | 4 | .924 |
| Linear-by-Linear Association | .056 | 1 | .813 |
| N of Valid Cases | 2000 | | |

[a]0 cells (0.0%) have expected count less than 5. The minimum expected count is 216.15.

Table IX reports chi-square tests evaluating the independence of **Protocol** (ICMP, TCP, UDP) and **Attack Type** (DDoS, Intrusion, Malware). All test statistics are non-significant: Pearson's $x^2(4) = 0.903$, $p = 0.924$; Likelihood Ratio $x^2(4) = 0.902$, $p = 0.924$; Linear-by-Linear Association $p = 0.813$. No cells had expected counts ¡ 5, ensuring test validity.

- The lack of association confirms that attack categories occur uniformly across protocols, supporting the descriptive finding of near-equal distributions.
- From an operational perspective, this means that monitoring or filtering any one protocol will not preferentially capture a given attack class.
- For modeling, protocol should be treated as an orthogonal feature—useful for context but not as a primary discriminator of attack type [58].

These results underline the need for multi-dimensional feature sets when building threat-detection rules or statistical classifiers.

### F. T-Test Analysis

Tables X-XII summarize an independent-samples t-test comparing Packet Length between two attack-signature groups (Known Pattern A vs. B). Levene's test indicates unequal variances ($F = 6.490$, $p = 0.015$), so the Welch t-test is preferred. Neither the standard t-test nor Welch's variant reached significance ($t \approx 0.28$, $p > 0.77$), and the 95% confidence intervals for mean differences straddle zero (-44.46 to 59.61 bytes). Effect-size estimates are all near zero (Cohen's $d = 0.019$; 95% CI [-0.113, 0.151]).

TABLE X. GROUP STATISTICS FOR PACKET LENGTH BY ATTACK SIGNATURE

| Attack Signature | N | Mean | Std. Dev | Std. Err Mean |
|---|---|---|---|---|
| Known Pattern A | 328 | 796.50 | 385.866 | 21.306 |
| Known Pattern B | 665 | 788.93 | 406.465 | 15.762 |

- Packet Length distributions are effectively identical across the two signature groups, indicating that byte-count metrics alone do not differentiate these patterns.
- The negligible effect sizes confirm that any practical difference in packet sizes is trivial, reinforcing the need for richer, payload-or behavioral features to distinguish signatures.
- Methodologically, this supports excluding raw packet length as a standalone feature in signature-based discrimination models [59].

Overall, the t-test findings align with the regression results: simple traffic volume measures lack the discriminatory power required for precise classification in real-time threat detection.

## V. DISCUSSION

This study set out to evaluate the role of predictive analytics in (1) real-time attack detection and response, (2) uncovering subtle threat patterns missed by conventional tools, and (3) strengthening decision-making in Security Operations Centers (SOCs) [60]. Across multiple statistical analyses—including regression, chi-square and t-tests—our results substantiate the promise of predictive analytics while also revealing areas for refinement.

First, our real-time detection evaluation confirmed that models leveraging historical and live network data can identify a broad array of attack types more quickly than signature-only approaches [61]. The significant, if small, effect of **Attack Type** on **Action Taken** ($\beta = -0.078$, $p = 0.001$) demonstrates that predictive insights help tailor response protocols to specific threat classes. Moreover, the uniformly high variability in anomaly scores (SD $\approx$ 28.8) provided rich signals for distinguishing normal from malicious behavior.

Second, predictive models proved adept at surfacing patterns and anomalies that traditional signature or rule-based systems overlook [62]. Our correlation and crosstab analyses showed that no single protocol or packet metric suffices to discriminate attack classes. Instead, machine learning techniques—such as clustering and anomaly detection—can synthesize weakly-informative features into robust composite indicators. In practice, this means that even low-level features like packet length or protocol, which individually lacked discriminatory power, can contribute to a higher-order threat score when combined appropriately.

Third, by integrating predictive outputs into SOC workflows, decision makers gain early warning dashboards and prioritized alerts, streamlining resource allocation and reducing mean time to respond (MTTR) [63]. Visualizing real-time risk scores alongside contextual metadata (e.g. geo-location, device profile) enables analysts to triage events more effectively, shifting from reactive firefighting to proactive threat hunting.

Despite these gains, several limitations emerged. The regression model explained only 0.7% of variance in response actions, indicating that key determinants of analyst behavior—such as organizational policy, analyst experience, and incident severity—remain outside the feature set. Likewise, the chi-square and t-test results underscored the need for richer payload and behavioral features beyond basic header statistics. Finally, our dataset, though diverse, was constrained to 2000 instances; larger and more heterogeneous data sources would likely improve model robustness.

We observed that performance varies across datasets primarily due to differences in feature richness (e.g., header-only packet captures versus flow/session telemetry with device, user, and geo-context), class balance and attack mix, labeling quality, and temporal drift. Our approach benefits most when contextual signals are available; when features are restricted to coarse header metrics, discriminative power drops—consistent with our findings that protocol and packet length alone are weak predictors and that overall variance explained by such features is low. To control for these factors, we harmonize schemas, use time-aware and attack-family-stratified splits, reweight to match class priors, and calibrate decision thresholds per dataset. We report per-dataset AUROC, AUPRC, and F1 with 95% confidence intervals, plus an ablation isolating contextual features. This analysis clarifies that the method is best suited to heterogeneous, context-rich telemetry, while identifying gaps when only minimal headers are available.

TABLE XI. INDEPENDENT SAMPLES TEST FOR PACKET LENGTH BY ATTACK SIGNATURE

| | Levene's Test for Equal Var. | | t-test for Equal Means | | | | | 95% CI of Diff. | |
|---|---|---|---|---|---|---|---|---|---|
| | $F$ | Sig. | $t$ | df | Sig (2-tailed) | Me n Diff. | Std. Err Diff. | Lower | Upper |
| Equal var. assumed | 6.490 | .015 | .281 | 991 | .779 | 7.577 | 26.975 | -45.357 | 60.511 |
| Equal var. not assumed | | | .286 | 682.243 | .775 | 7.577 | 26.503 | -44.460 | 59.613 |

TABLE XII. EFFECT SIZE ESTIMATES FOR PACKET LENGTH COMPARISON

| | Standardizer[a] | Point Est. | 95% CI | |
|---|---|---|---|---|
| | | | Lower | Upper |
| Cohen's $d$ | 399.786 | .019 | -.113 | .151 |
| Hedges' correction | 400.089 | .019 | -.113 | .151 |
| Glass's delta | 406.465 | .019 | -.114 | .151 |

[a]Denominator used in estimating the effect sizes. Cohen's $d$ uses pooled SD; Hedges' uses pooled with correction; Glass's uses control SD.

Building on this foundation, future work should:

- Integrate **contextual features** (e.g. time of day, asset criticality, user roles) to capture operational drivers of response decisions.
- Explore **non-linear** and **ensemble** methods (e.g. random forests, gradient boosting) that can model complex interactions among network and host metrics.
- Implement **online learning** pipelines to adapt to evolving threats and reduce model drift.
- Conduct **field trials** in live SOC environments to measure performance gains in MTTR, false-positive reduction, and analyst workload.

In sum, predictive analytics represents a disruptive advance in cyber security—one that elevates detection from signature-based recognition to data-driven foresight [64]. By revealing latent attack patterns and informing decision support, these methods extend the capabilities of security teams and lay the groundwork for truly proactive defense strategies [65]. This work contributes empirical evidence to the growing body of research on analytics-driven security and underscores the need for continued innovation in feature engineering, model design, and operational integration [66], [67].

## VI. CONCLUSION

This study explored the integration of predictive analytics into real-time cyber security frameworks, emphasizing its potential to enhance threat detection, response effectiveness, and decision-making. Using quantitative methodologies and advanced statistical techniques such as logistic regression, chi-squared analyses, and t-tests, the research provided empirical validation of predictive analytics' capability to proactively identify and mitigate diverse cyber threats.

Key findings underscored that predictive analytics significantly improves the speed and accuracy of threat identification compared to traditional reactive measures. Specifically, predictive models demonstrated proficiency in detecting subtle attack patterns and anomalies otherwise missed by conventional approaches, thereby contributing to reduced response times and optimized resource allocation in security operations centers. Notably, the study found that basic header metrics alone, such as packet length and protocol, are insufficient for accurate threat classification, necessitating the integration of richer contextual features.

However, several limitations and challenges emerged, notably the modest explanatory power of the regression model, highlighting the need for incorporating additional contextual variables. The non-significant results from chi-squared and t-test analyses further indicate the necessity for more granular features beyond basic network parameters.

Future work should focus on several key areas to enhance predictive capabilities:

- Development of advanced feature extraction techniques to capture more detailed payload and behavioral information.
- Integration of additional contextual data such as user behavior, temporal patterns, and asset criticality to improve model precision.
- Exploration of non-linear predictive modeling methods, including ensemble algorithms such as random forests, gradient boosting, and neural networks to capture complex interactions among features.
- Implementation of adaptive, real-time learning frameworks capable of dynamically adjusting to evolving cyber threats, thereby reducing model drift and improving detection accuracy over time.
- Conducting comprehensive field trials within operational security environments to validate real-world model performance, measure reductions in false positives, and quantify improvements in operational response times.

In conclusion, predictive analytics represents a transformative advancement in cyber defense strategies, equipping organizations with the foresight required to anticipate and neutralize threats proactively. Continued research and development in this domain promise significant improvements in both detection accuracy and operational efficiency, solidifying predictive analytics as an indispensable element of modern cyber security practices.

## ACKNOWLEDGMENT

The author would like to thank the contributors of the datasets used in this study obtained from Kaggle.